# Uniformity of response of Uranium fission chambers used as neutron beam monitors

**F. Lafont**[a1], **B. Guerard** [a], **R. Hall-Wilton**[b] **and K. Kanaki**[b]

[a] *Institut Laue Langevin,*
 *71 Avenue des martyrs 38042 Grenoble, France*
[b] *European Spallation Source ERIC,*
 *Box 176, SE-221 00, Lund, Sweden*
 *E-mail:* `lafont@ill.fr`

ABSTRACT: Uranium fission chambers are commonly used for neutron beam monitoring in neutron research facilities. The challenges brought with the renewing or commissioning of new neutron facilities encourage to minimize the interaction between the monitor and the neutron beam. In order to characterize both their detection efficiency and transmission factor, several fission chambers, have been scanned on the CT1 monochromatic beam line of the ILL. Results show that the monitor transparency, measured at the center of the monitor, varies from 90% to 96%, and that the detection efficiency varies by as much as 50% over the active area.

---

[1] Corresponding author.

**Content**

## 1. Introduction

Transmission beam monitors are used on many neutron instruments to measure the neutron beam intensity and its variation in order to calibrate experimental data. Uranium fission chambers have demonstrated their reliability as well as several advantages compared to other technologies: thanks to the high Q-value of the fission reaction, no gas amplification is required and the energy deposited in the gas is several orders of magnitude higher for neutron events than for gamma events. As a result, the gamma discrimination is excellent, and very intense neutron beams are accessible. Furthermore, the absence of gas amplification provides immunity against counting variations due to temperature changes, or high voltage instabilities. More than 15 instruments of ILL are equipped with one or several fission chambers with detection efficiencies ranging from $10^{-3}$ to $10^{-7}$. Nonetheless, they still have some drawbacks:
- fission interactions produce fast neutrons that might contribute to background,
- Uranium neutron absorption probability is not linear in the 1 Å neutron wavelength region which makes difficult the comparison between measurements at different wavelengths in this range,
- The materials, from which fission chambers are made, induce some beam losses.

The monitor transparency is defined as the proportion of neutrons passing through the monitor without interacting with its materials. This paper is intended to measure this critical parameter as well as the uniformity of the neutron detection efficiency. The results are discussed under the light of the limited information we have got concerning the fabrication of these devices.

## 2. Transparency and efficiency measurement

### 2.1 Measurement setup

The measurement was performed on the monochromatic (2.5 Å) CT1 test beamline of the ILL. The beam is collimated by $B_4C$ absorber plates; its section is 3 mm × 3 mm. The monitors to be tested are mounted behind this neutron-absorbing collimation on a dual-axis translation table; this allows moving the monitor in the plane perpendicular to the beam axis. A full absorption $^3$He counter, called reference detector, is located 1 m away to measure the transmitted neutron flux. Different models of Uranium monitors all coming from LND Inc [1], covering different efficiency ranges, were tested.

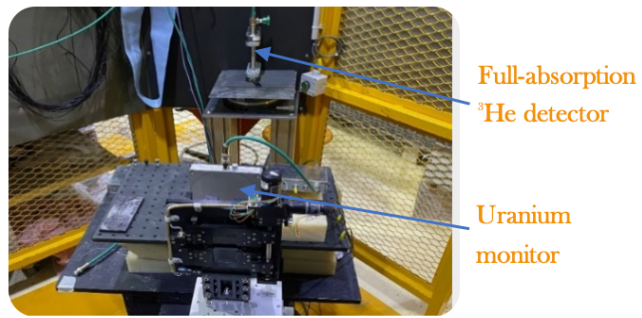

**Figure 1**. Transparency measurement setup on CT1 at ILL

Beam monitors with model numbers 3053, 3054 and 30521 (custom model) have theoretical efficiencies at 1.8 Å of $10^{-3}$, $10^{-4}$ and $10^{-5}$ respectively. The sensitive areas are made of Uranium compound coated disks of 101.6 mm diameter for the 3053 and 3054 models, and 128.8 mm for the 30521. A positive voltage is applied to those disks (anode) that are encapsulated in a sealed Aluminum chamber (cathode) filled with Argon and Nitrogen. The fission products from the neutron interactions with Uranium transfer their energy to the gas by ionization; the created charge is collected on the anode. After amplification, a discriminator circuit selects signals whose amplitude is high enough to be counted as a neutron event.

These three beam monitors were scanned on their full surface. For each position of the monitor, we measured the counting rate of both the beam monitor, and the reference detector. For ten other beam monitors with efficiencies ranging from $10^{-3}$ to $10^{-7}$ we did the same measurement with a 1 cm$^2$ beam, but only at the center of the beam monitor. Apart from the beam size, the setup remains unchanged and only the Uranium monitor is switched from one measurement to another. Since most of the scattered neutrons are lost and not counted by the reference detector, the definition to the transparency to neutrons combines losses from absorption and scattering.

**2.2 Monitor transparency measurement**

The results of the transparency measurements in Figure 2 show that the neutron beam attenuation varies between 2.5% to 7.5% over the active area. This variability most likely comes from the coating process and is discussed in Section 3. For the 3053 and the 3054, we observe a lower transparency in the middle of the sensitive area compared to the borders.

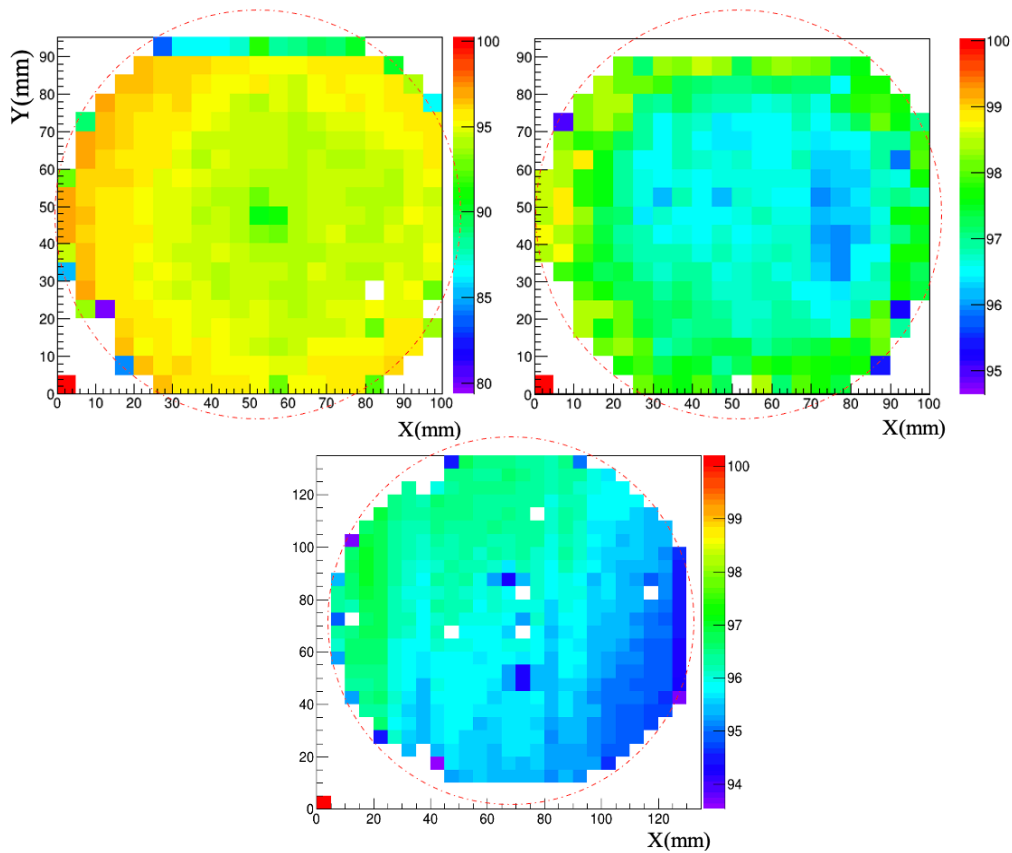

**Figure 2**. Transparency measurement of monitors 3053 (top left), 3054 (top right) and 30521 (bottom). The dashed red circle indicates the theoretical sensitive area of each monitor. The color bar corresponds to the neutron transparency in %.

Figure 3 shows the measured neutron transparency as a function of the Uranium SMD (Surface Mass Density) as indicated by the manufacturer. The statistical error is limited to less than 0.1 % as well as the systematics. Indeed, the fraction of detected scattered neutrons is less than 0.1%. There is no correlation between transparency and Uranium SMD, and we observe a large distribution of the transparency values for a given Uranium SMD. The beam attenuation due to neutron absorption in Uranium is less than 1 ‰. Considering that the measured attenuation is far greater, most of the neutron interactions take part with the other materials contained in the beam monitor, namely the Aluminum windows, the substrate of the neutron converter film, and the molecules contained in the converter film apart from the Uranium.

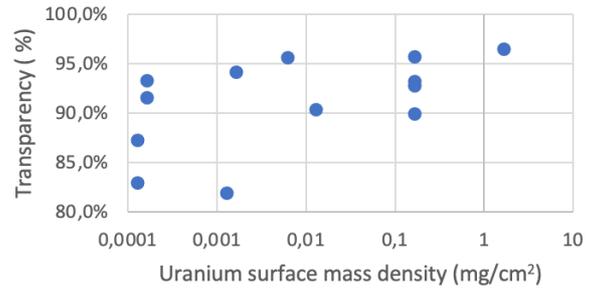

**Figure 3**. Transparency at the center of several beam monitors depending on the Uranium surface mass density. Errors bars (~0.1 %) are too small to be visible at this scale.

### 2.3 Neutron detection efficiency

Figure 4 shows the efficiency mapping for the same 3053, 3054 and 30521 models as in Figure 2. For each pixel, the detection efficiency is given by the counting rate of the monitor divided by the counting rate of the reference detector.

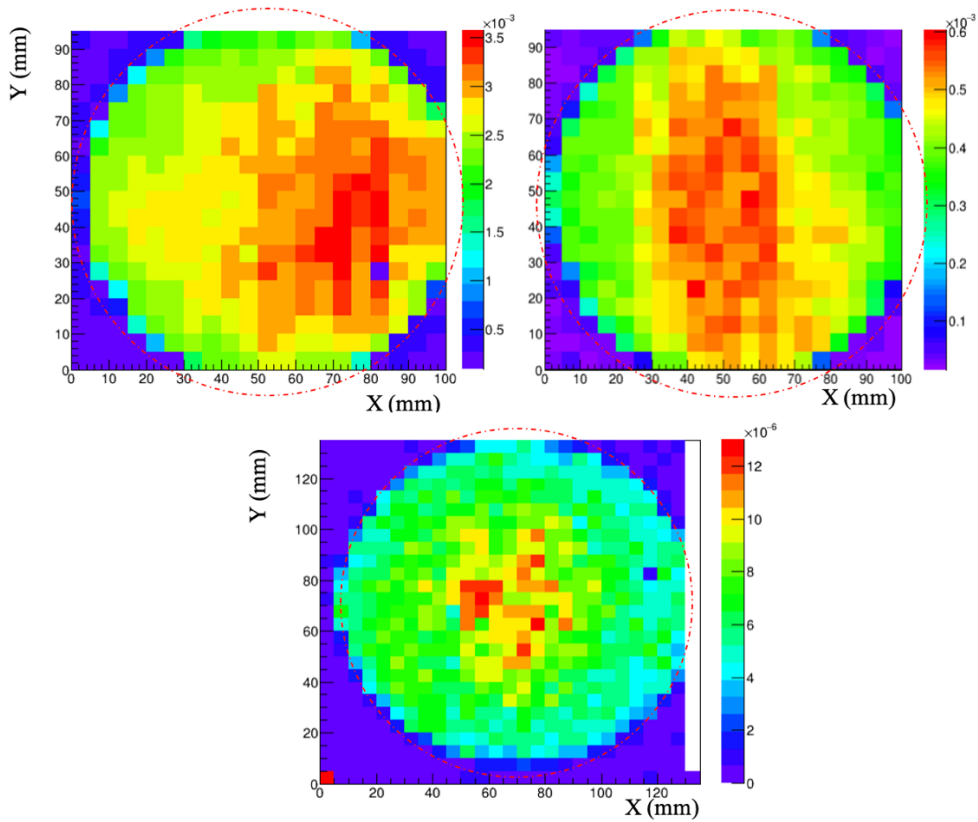

**Figure 4**. Uniformity of response of U-monitors 3053 (top left) and 3054 (top right) and 30521 (bottom). The dashed red circle indicates the theoretical sensitive area of each monitor. The color bar corresponds to the absolute efficiency values.

For the three monitors, we see a high efficiency zone covering between 30% and 50% of the sensitive area. The measured deviations from the mean efficiency are ±15 %, ± 25%, and ± 50% for the 3053, 3054, and 30521 model respectively. The non-uniformity gets worse when the detection efficiency is reduced, and it is even beyond the supplier specifications (± 30%) for the 30521. Also, if we define the active area limits as a drop of 50 % of the mean efficiency value to account for the 3x3 mm beam cross section, the interpolated diameter of the active area is 100 $^{+2}_{-5}$ mm instead of the 101.6 mm as specified for the 3053 and 3054 models and 124 $^{+2}_{-5}$ $mm$ instead of the 128.7 mm for the 30521.

Comparing these results with those in the previous section, we see that, for each monitor, the transparency decreases while the efficiency increases. The relationship between efficiency and transparency measured on the different points of one monitor is given in Figure 5. It shows a clear correlation for the 3053 and no correlation at all for the 30521. There is some correlation for the 3054, but not as good as for the 3053. For these three monitors, the higher the efficiency is, and the better the correlation is between transparency and efficiency.

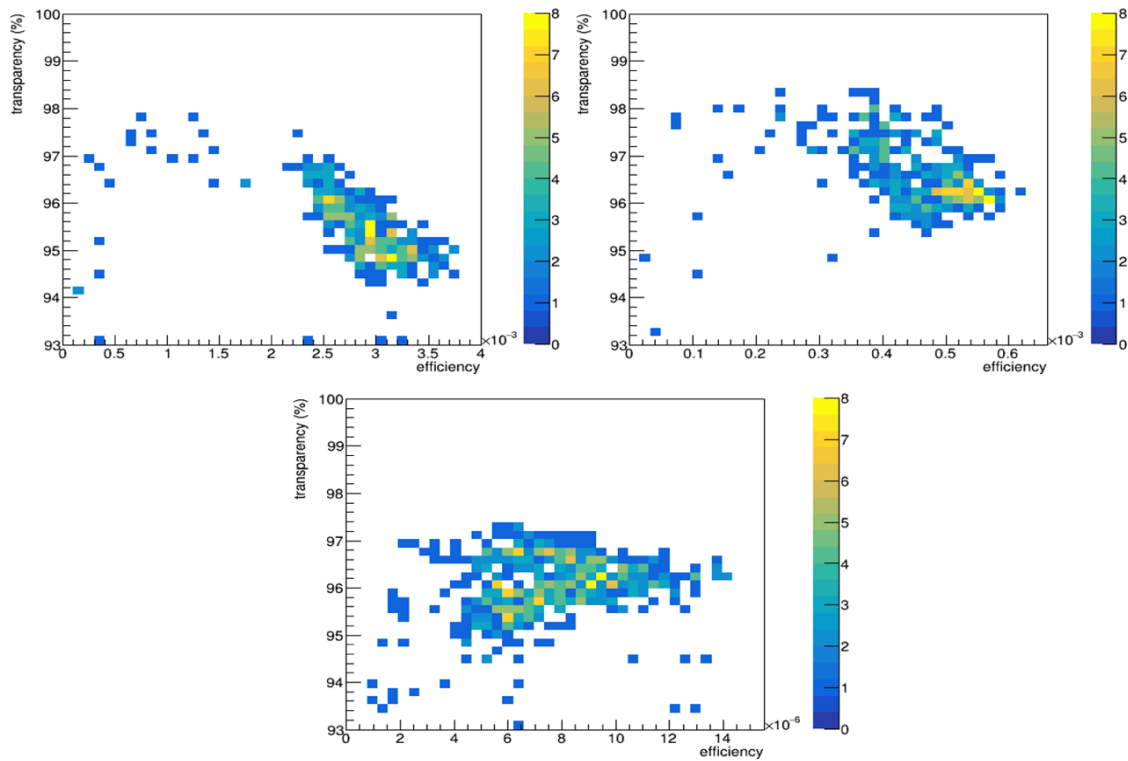

**Figure 5**. Scatter plots of efficiency vs transparency for 3053 (top left), 3054 (top right) and 30521 (bottom) Uranium monitors.

Even if we did not scan all beam monitors on their whole surface, we observed the same behavior for all of them, namely a significant variation of the detection efficiency and of the transparency measured depending on which part of the monitor is exposed to the beam. Besides, we measured large differences (a factor of 2) in the counting rate for different beam monitors of the same model.

We also scanned the pencil-like fission chambers (models 30765, 30766, 30767 with respectively $10^{-5}$, $10^{-6}$ and $10^{-7}$ theoretical efficiencies …). These cylindrical beam monitors have an internal diameter of 10 mm, and their theoretical sensitive length is 63 mm. Thanks to their compactness, they can be mounted in tiny spaces. As can be seen in Figure 6, we measured a sensitive length shorter than the one specified by the manufacturer, and a very non-uniform detection efficiency along the sensitive length.

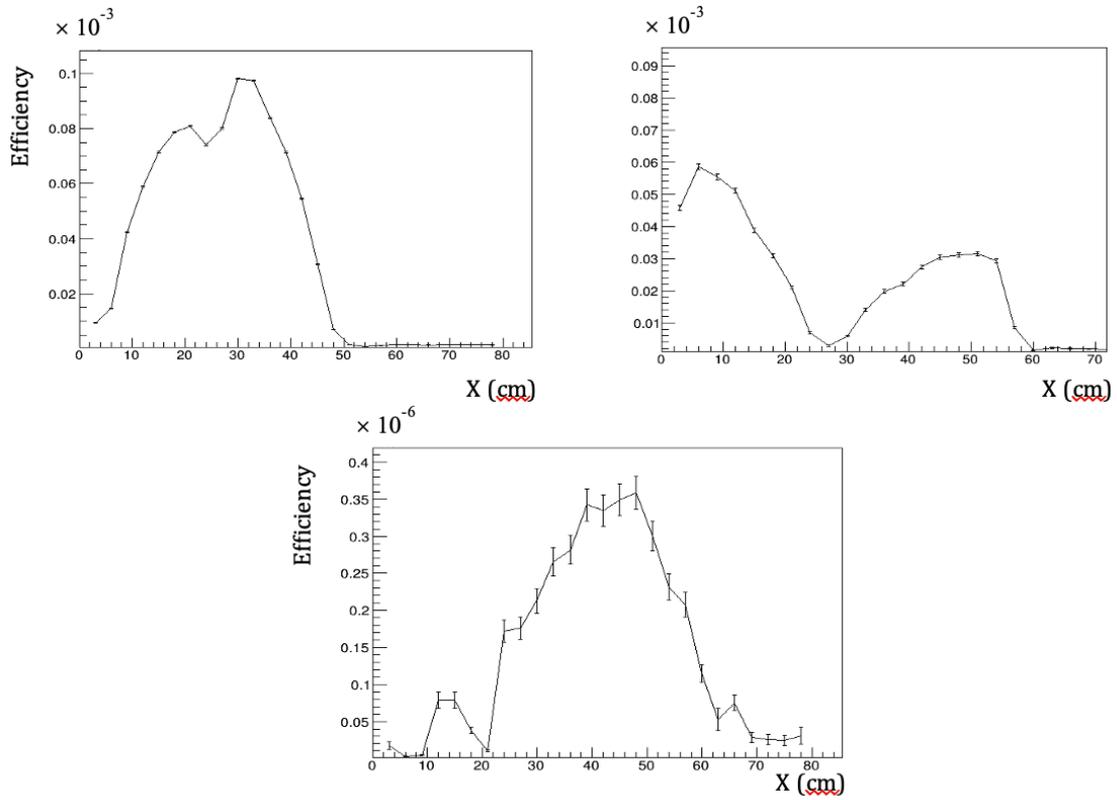

**Figure 6**. Detection efficiency measured along several pencil-like Uranium monitors.

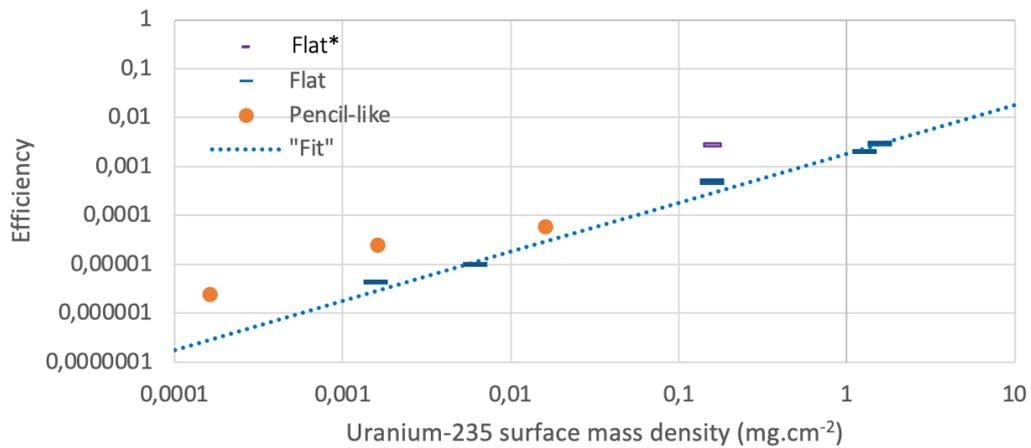

**Figure 7.** Efficiency of several Uranium monitors depending on Uranium surface mass density. Errors bars are too small (<1%) to be visible at this scale.

Figure 7 shows the relationship between the Uranium SMD and the measured detection efficiency. For the three pencil-like monitors, the efficiency is taken as the average value in Figure 6, whereas for the flat monitors it is the efficiency measured at the center. We fit the seven flat fission chambers data points with equation (1).

$$eff = 1 - e^{\frac{-\sigma N_a \mu}{M}}, \qquad (1)$$

where *eff* is the efficiency, $N_a$ is the Avogadro constant, M is the $^{235}$U compound molar mass, µ is the Uranium surface mass density and σ is the $^{235}$U fission cross section. One of the data points (Flat*), corresponding to a 3054 model, was excluded from the fit because the LND reference is probably wrong since its efficiency corresponds most likely to a 3053 model. The results from the fit gives a value of σ equal to 797±31 b, which matches well the theoretical cross section of 835 b [2] for the 2.5 Å wavelength of CT1. This means that the $^{235}$U SMD is generally correct to achieve the specified efficiency at 1.8 Å, but we measured deviations up to 75 % for the monitors tested.

## 3. Discussion

The measurements performed with 14 flat fission chambers produced by LND demonstrate significant non-uniformities over the active area for both the transparency and the neutron detection efficiency. Non-uniformities are even higher for the three pencil-like fission chambers tested.

The flat fission chambers contain two windows of 1 mm thickness each, and an internal Aluminum plate of 0.6 mm on which the neutron converter film is electrodeposited. The electrodeposition was performed by Eckert and Ziegler company. We assume that the transparency of the Aluminum windows and of the internal plate are uniform. We have seen that the Uranium contributes only marginally to neutron interactions; hence, we can deduce that most of the transparency non-uniformity is due to variations of the SMD of the molecules associated to Uranium inside the converter film. We also observed a correlation between transparency and detection efficiency for the 3053-type fission chamber which has the higher $^{235}$U SMD, and that this correlation degrades when $^{235}$U SMD decreases (models 3054 and 30521). This could be explained by the fact that the lower the $^{235}$U SMD, the higher the statistical fluctuation, i.e. the fluctuation of detection efficiency.

Furthermore, the absence of correlation between the mass of Uranium and the transparency of the detector (Figure 3) suggests that the density of U-particles in the electrodeposition mixture is the parameter used by Eckert and Ziegler to get the Uranium SMD required. We know that Uranium particles are electrodeposited on 0.3 mm-thick Aluminum plates, and that two plates are mounted in the middle of the fission chamber, with the 2 Uranium films on opposite sides [3], but we do not know what the Uranium films are made of, apart from Uranium.

In order to complete our measurements, X-ray and neutron transmission images of the 3053-model were made (Figure 8) on the D50 instrument [4] at the ILL with flat field and background corrections. The "neutron image" should not be directly compared to the image of Figure 2 since the neutron wavelengths are different for the 2 instruments. The fact that the beam monitor was mounted close to the D50 camera can explain the apparent higher transparency since a fraction of the scattered neutrons might be detected. The most interesting information we got from the neutron and X-Ray transmission images is the singularity in the center of the monitor: there are dots that absorb both neutron and X-rays. This defect is located on both plates, which suggests that this is inherent to the electrodeposition process. Although we do not know the origin, these spots do not seem to correspond to a higher surface density of Uranium since the neutron detection efficiency is not different from the rest of the fission chamber. This assumption is correct considering that the absorption length of the fission products inside the converter film remains larger than the film thickness. Due to the lack of information from the supplier, we can only suspect that it comes from unexpected heavy metals – possibly from the anode – deposition. A better knowledge of the electrodeposition process and of the film composition and thickness would be necessary to fully understand these results.

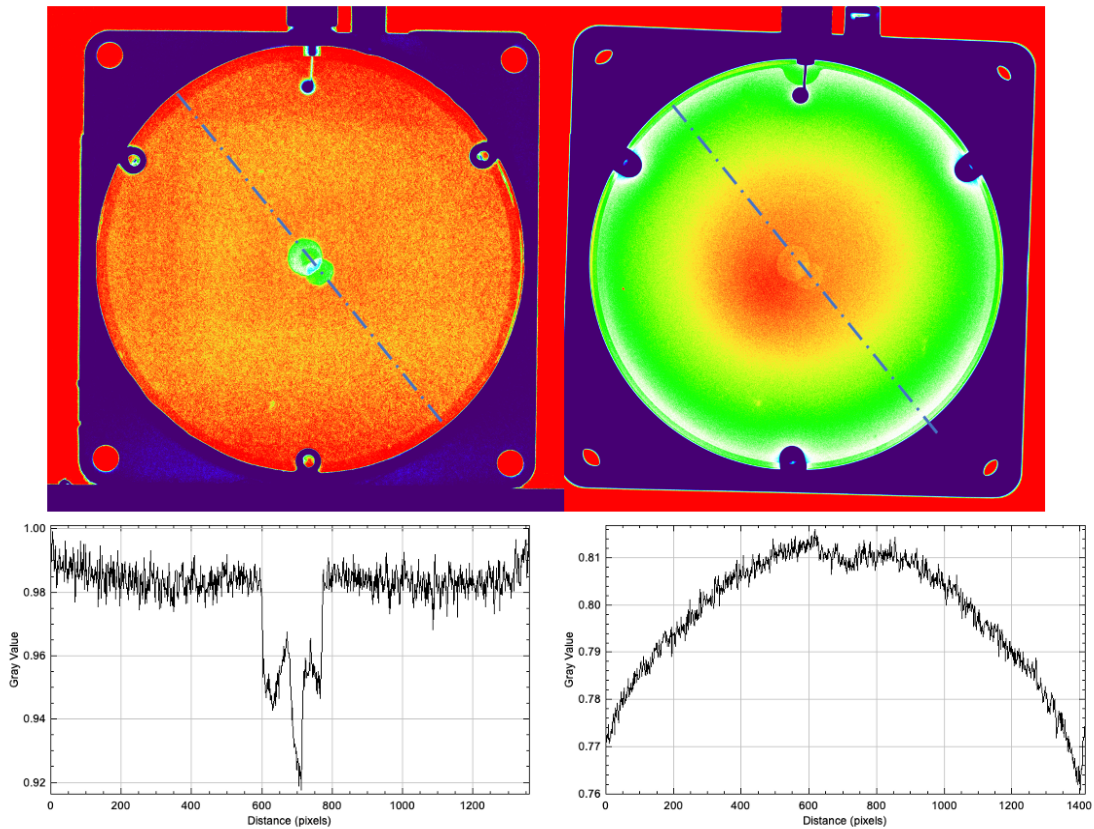

**Figure 8.** Neutron (left) and X-ray (right) radiographies of the 3053 Uranium monitor with a respective profile plot in grey values along a diagonal (dotted line) of the detection area.

## 4. Conclusion

Uranium neutron beam monitors have interesting characteristics and had proven their reliability over time. Nonetheless, the results presented here show that they have inherent drawbacks, which, to our knowledge, were not considered before. From our measurements, we observed non-uniformities of both the transparency and the detection efficiency that most likely come from the Uranium electrodeposition process. This could be an issue if data acquired in different beam conditions have to be normalized to correct variations of the beam intensity, for example if different collimations or wavelengths of the beam are used during one experiment. More critical is the discrepancy observed in several cases, sometimes by more than one order of magnitude, between the specified detection efficiency and the measured one. Finally beam losses due to neutron scattering in the converter film is responsible for a reduction of the beam intensity on one side, and for a higher radiation background resulting in a thicker radiation shielding needed on the other side, at least when the monitor is located close to the sample. For in-bunker operation, this type of monitor can still be a high-rate capable and reliable solution assuming the instrument can tolerate the reduction of the beam flux.

These aspects combined with the fact that some instruments cannot afford to have fast neutrons produced by Uranium fissions is motivation to develop new-generation monitors that would use other neutron converters and would not suffer from these problems.

## Acknowledgments

This work was carried out at the ILL, in a collaborative effort to find out the best options for the neutron beam monitoring at ESS and ILL facilities.